\documentclass[floats,floatfix,prl,amssymb,twocolumn,superscriptaddress,nofootinbib,10pt]{revtex4-1}

\usepackage{amssymb,amsmath,verbatim,mathtools,needspace,enumitem,etoolbox,graphicx,physics,microtype,afterpage,bm}
\usepackage[dvipsnames, usenames]{xcolor}
\definecolor{linkcolor}{rgb}{0.0,0.3,0.5}
\usepackage[unicode, colorlinks=true, linkcolor=linkcolor, citecolor=linkcolor, filecolor=linkcolor,urlcolor=linkcolor, pdfusetitle]{hyperref}
\usepackage[all]{hypcap}
\usepackage[T1]{fontenc}
\usepackage[utf8]{inputenc}
\usepackage{tabularx}
\usepackage{float}
\usepackage{lmodern}
\allowdisplaybreaks
\usepackage{tikz}
\usepackage{color}
\usepackage{framed}
\usepackage{hyperref}
\hypersetup{colorlinks, citecolor=bluscuro, linkcolor=black, urlcolor=bluscuro}
\definecolor{rossos}{cmyk}{0,1,1,0.55}
\definecolor{bluscuro}{rgb}{0.15, 0.2, .85}
\definecolor{bluchiaro}{cmyk}{1,.3,0.,0.1}
\definecolor{ForestGreen}{rgb}{0.13, 0.55, 0.13}

\newcommand{\be}{\begin{equation}}
\newcommand{\ee}{\end{equation}}
\newcommand{\bea}{\begin{equation}\begin{aligned}}
\newcommand{\eea}{\end{aligned}\end{equation}}
\renewcommand{\d}{{\rm d}}

\def\lsim{\mathrel{\rlap{\lower4pt\hbox{\hskip0.5pt$\sim$}}
    \raise1pt\hbox{$<$}}}         
\def\gsim{\mathrel{\rlap{\lower4pt\hbox{\hskip0.5pt$\sim$}}
    \raise1pt\hbox{$>$}}}         

\def\g{h}
\def\d{{\rm d}}

\def\o{m}

\begin{document}

\title{Explaining  Nonlinearities in Black Hole Ringdowns from Symmetries}

\author{Alex Kehagias}
\email{kehagias@central.ntua.gr}
\affiliation{Physics department, National Tecnical University of Athens, 15780 Athens, Greece}
\affiliation{CERN, Theoretical Physics Department, Geneva, Switzerland}

\author{Davide Perrone}
\email{davide.perrone@unige.ch}
\affiliation{D\'epartement de Physique Th\'eorique, Universit\'e de Gen\`eve, 24 quai E. Ansermet, CH-1211 Geneva, Switzerland}

\author{Antonio Riotto}
\email{antonio.riotto@unige.ch}
\affiliation{D\'epartement de Physique Th\'eorique, Universit\'e de Gen\`eve, 24 quai E. Ansermet, CH-1211 Geneva, Switzerland}
\affiliation{ Gravitational Wave Science Center (GWSC), Universit\'e de Gen\`eve, 24 quai E. Ansermet, CH-1211 Geneva, Switzerland}

\author{Francesco Riva}
\email{francesco.riva@unige.ch}
\affiliation{D\'epartement de Physique Th\'eorique, Universit\'e de Gen\`eve, 24 quai E. Ansermet, CH-1211 Geneva, Switzerland}


\begin{abstract}
\noindent
It has been recently pointed out that nonlinear  effects are necessary to  model the ringdown stage of 
the gravitational waveform produced by the merger of two  black holes giving rise to a remnant Kerr black hole.
We show that  this nonlinear behaviour is  explained,  both on the qualitative
and quantitative level,  by near-horizon symmetries of the Kerr black hole within the Kerr/CFT correspondence.
\end{abstract}

\maketitle

\noindent{{\bf{\it Introduction.}}}
 Quasi Normal Modes (QNMs) provide a unique tool  to investigate the properties of Black Holes (BHs) whose understanding is one of the   major goals of gravitational wave astronomy \cite{Berti:2009kk}.  
 In the merger of two BHs, during the  final stage called ringdown,  they  dominate the BH  response to any kind of disturbance and their frequencies are uniquely determined by the BH mass, spin and charge. 

Gravitational Waves (GWs) during ringdown are well   described by a superposition of exponentially damped QNMs, labeled by two angular harmonic numbers $(\ell,m)$ and an overtone number $n$. Their amplitude is denoted $A_{(\ell,m,n)}$, while their oscillation frequency and decay timescale are given by  the real and imaginary parts of~$\omega_{(\ell,m,n)}$. The GW strain far from the BH source can be decomposed as

\begin{eqnarray}
h(u,\theta,\phi)&=&
\sum_{\ell\geq 2}\sum_{|m|\leq \ell} h_{(\ell,m)}(u)
{}_{-2}Y_{(\ell, m)}(\theta,\phi),\nonumber\\
h_{(\ell,m)}(u)&=&\sum_{n\geq 0}A_{(\ell,m,n)} e^{-i\omega_{(\ell,m,n)}(u-u_{\text{\tiny pk}})},
\end{eqnarray}
where  $u=(t-r)$ is  the Bondi time, $u_{\text{\tiny pk}}$
 is the time at which the strain achieves its maximum value, 
and ${}_{-2}Y_{(\ell, m)}$ are  the spin-weight $s= -2$ spherical harmonics (the $s=+2$ mode with outgoing boundary conditions is subleading at infinity \cite{Teukolsky:1973ha}). 
The  GW strain produced   is generically  modeled using first-order BH perturbation theory. However, nonlinearities are  an intrinsic property of general relativity and indeed 
it has been recently pointed out that   second-order
effects are relevant to describe  ringdowns from BH merger simulations \cite{nl3,nl2} (see also Refs. \cite{nlold1,nlold2,Lagos:2022otp}). 

In particular, the
second-order  mode  amplitude $A^{(2,2,0)\times (2,2,0)}_{(4, 4)}$ obtained from the square of the first-order fundamental mode $(\ell,m)=(2, 2)$ 
is comparable to or even larger than that of the fundamental linear mode $(4,4)$. For the numerical simulations of quasicircular mergers giving rise to a Kerr BH with  spin $0.7$  (in units of the BH mass and we set from now on $G_N=1$) Ref. \cite{nl3} found   

\begin{equation}
\label{h}
\frac{\left|A^{(2,2,0)\times (2,2,0)}_{(4, 4)}\right|}{\left|A_{(2, 2,0)}\right|^2}=0.1637 \pm 0.0018,
\end{equation}
where  we have neglected the milder dependence on the BH mass ratio of the two BH mergers. This result is consistent with what found in Ref. \cite{nl2} which quotes values in the interval $(0.15-0.2)$.

Restricting ourselves to the fundamental modes and noting that the  second-order QNM $(4, 4)$ is sourced at second-order  from the square of the $(2, 2)$ mode with  frequency $2\omega_{(2, 2,0)}$,  Eq. (\ref{h}) can be written in the  suggestive form 

\begin{equation}
\label{three}
\frac{\langle
h_{(2, 2)}h_{(2, 2)}
 h_{(4, 4)}\rangle}{\langle h^2_{(2, 2)}\rangle^2 }
\simeq 0.1637 \pm 0.0018.
\end{equation}
Similarly, Ref. \cite{nl3} found

\begin{equation}
\label{h1}
\frac{\left|A^{(2,2,0)\times (3,3,0)}_{(5, 5)}\right|}{\left|A_{(2, 2,0)}\right|\left|A_{(3, 3,0)}\right|}=
   0.4735 \pm 0.0062,
\end{equation}
again for a Kerr BH remnant of spin $\sim 0.7$.
In order to  correctly model the BH ringdown one needs therefore to include  nonlinear effects. 

The goal of this paper is to explain the nonlinearities of the Kerr BH remnant from symmetry arguments. Our starting point is the realization that the QNMs are produced in the proximity of  the BH horizon \cite{Maggiore:2018sht}. In the region very close to the horizon of an extreme Kerr BH one can 
set  consistent boundary conditions such that  the asymptotic symmetry generators identify one copy of the Virasoro algebra. 
This implies that the near-horizon quantum states can be identified with those of (a
chiral half of) a two-dimensional conformal field theory (CFT) with finite temperature $T=1/2\pi $.  This goes under the name of  the Kerr/CFT correspondence \cite{Guica:2008mu}. Although the CFT details are not exactly known, several nontrivial checks have been studied in the past (see, for instance, Refs. \cite{c1,c2,c3}), consolidating the idea that there is a relation between universal properties of BHs and CFTs, and trying also to extend the results to non-extremal cases \cite{b1,b2,b3,Becker:2010jj}. 

If the correspondence is correct,   the correlators of  bulk fields  can be computed through the correlators of the corresponding boundary  operators. In turn, the latter    are dictated by the CFT. In the case of the spin-2 strain, the associated  boundary operator is the two-dimensional energy momentum tensor, whose correlator amplitudes are fixed by the central charge. This simple reasoning will allow us to calculate the three-point correlators  to estimate the level of nonlinearities in the ringdowns from symmetry arguments.

Before launching ourselves in the technicalities, we briefly summarise the basics of the Kerr/CFT correspondence. The expert reader on the subject can skip this part.

\vskip 0.5cm
\noindent{{\bf{\it Kerr/CFT.}}}
The Kerr BH with mass $M$ and angular momentum $J$ is described  in Boyer-Lindquist coordinates by the metric 

\begin{eqnarray}
    \dd s^2 &=& -\frac{\Delta}{\rho^2}\left( \dd \hat t - a \sin^2 \theta \right)^2 + \frac{\rho^2}{\Delta}\dd \hat{r}^2 + \nonumber\\
    &+&\frac{\sin^2 \theta}{\rho^2} \left[ (\hat{r}^2+a^2) \dd \hat \phi - a \dd \hat{t} \right]^2 + \rho^2 \dd \theta^2,
\end{eqnarray}
where
\begin{equation}
    \Delta = \hat{r}^2 -2M \hat{r} + a^2, \quad \rho^2 = \hat{r}^2 + a^2 \cos^2 \theta, \quad a=\frac{J}{M}.
\end{equation}
The two horizons are defined as the solutions to $\Delta(r) = 0$:
\begin{equation}
    \hat{r}_{\pm} = M \pm \sqrt{M^2 -a^2},
\end{equation}
and the Hawking temperature can be written as
\begin{equation}
    T_H = \frac{\hat r_+ - \hat r_-}{8 \pi M \hat r_+}\,.
\end{equation}
We  consider the extremal case $a=M$ and  the change of coordinates \cite{Bardeen:1999px}
\begin{equation}
    t = \frac{\lambda \hat{t}}{2M}, \quad r = \frac{ \hat{r}-M}{\lambda M}, \quad \phi = \hat{\phi} - \frac{\hat{t}}{2M},
\end{equation}
such that, in  the limit $\lambda \to 0$ and  keeping fixed the coordinates $(t, r, \phi, \theta)$, one can  zoom into a small region around the BH  event horizon $\hat r=M$. The resulting metric is the Near Horizon Extreme Kerr (NHEK)
\begin{eqnarray}
\label{metric}
    \dd s^2\hspace{-0.1cm} &=& \hspace{-0.1cm} 2 J \Gamma(\theta) \bigg[\hspace{-0.1cm} -r^2 \dd t^2 + \frac{\dd r^2}{r^2} + \dd \theta^2 + \Lambda^2(\theta) (\dd \phi + r \, \dd t)^2 \bigg],\nonumber\\
&&\Gamma(\theta) = \frac{1+ \cos^2 \theta}{2},\quad \Lambda(\theta) = \frac{2 \sin \theta}{1 + \cos^2 \theta},
\end{eqnarray}
with $\phi\sim\phi+2\pi$ and $0\leq \theta\leq \pi$. 
For a generic value  of $\theta$ one deals with  the geometry of a warped $\rm AdS_3$ with a ${\rm SL}(2, \mathbb{R})\otimes U(1)$ symmetry group.
At the specific  value  $\theta_0$ such that $\Lambda(\theta_0)=1$, the metric is that of gravity on ${\rm AdS}_3$ \cite{Anninos:2008fx, Bengtsson:2005zj}. 

At extremality, rotational perturbations along the angular azimuthal direction correspond  to excitations of the left sector of the dual CFT. The right modes are  not excited (unless one goes above extremality). On the other hand, the left modes cannot be associated with the Hawking temperature $T_H$, which vanishes for extremal Kerr BHs. 
Indeed, the Hartle–Hawking vacuum for quantum fields in the region outside the  Schwarzschild BH horizon (which gives rise to a density matrix $e^{-\omega/T_H}$) cannot be used for the Kerr spacetime. This is because in  spacetimes   lacking a  globally defined  timelike Killing vector, such as the Kerr geometry, the Hartle–Hawking vacuum does not exist. One can however  use the 
Frolov-Thorne vacuum  \cite{Frolov:1989jh}
which is appropriately defined in the vicinity of a spinning BH horizon.  
 Quantum fields  can be expanded in  asymptotic energy and angular momentum eigenstates  of the operators $\partial_{\hat{t}}$ and $\partial_{\hat{\phi}}$ as 

\begin{eqnarray}
\Phi_{s}(\hat{t},\hat{r},\theta,\hat{\phi})=\sum_{\omega,\ell,m}\Phi_{\omega\ell m}e^{-i\omega \hat{t}+i m\hat{\phi}} R_{\ell m}(\hat{r},\theta).
\end{eqnarray}
In the near horizon coordinates ($t,r,\theta,\phi$), the asymptotic energy-angular momentum eigenstates are expressed as 

\begin{eqnarray}
\label{ca}
e^{-i\omega \hat{t}+i m\hat{\phi}}=e^{-in_R t+i n_L\phi},
\end{eqnarray}
where 

\begin{eqnarray}
\label{cb}
n_L=m, \qquad n_R=\frac{1}{\lambda}\left(2 \omega M-m\right). 
\end{eqnarray}
Then the Frolov-Thorne vacuum gives rise to a density matrix with a Boltzmann weighting factor

\begin{eqnarray}
e^{-\frac{n_L}{T_L}-\frac{n_R}{T_R}}, 
\end{eqnarray}
where the left- and right-temperatures  
are \cite{Guica:2008mu}

\begin{eqnarray}
\label{TLTR}
T_L=\frac{r_+-M}{2\pi (r_+-a)}, \qquad T_R=\frac{r_+-M}{2\pi r_+\lambda}.
\end{eqnarray}
In the extremal  case (keeping $\lambda$ small, but finite), we end up with the  left-moving sector   in the Frolov-Thorne temperature 
\begin{eqnarray}
T_L=\frac{1}{2\pi},
\end{eqnarray}
and a Boltzmann suppression factor  $e^{-2\pi n_L}$, while the right sector has  $T_R=0$.
At the same time, the   geometry~(\ref{metric}) has a boundary at $r\to \infty$ (where boundary fields are a function of $t$ and $\phi$) and an asymptotic symmetry group (with  specific boundary conditions for the  perturbations of the metric in AdS) which extends the symmetry to half of a Virasoro algebra \cite{Guica:2008mu}. This result led to the Kerr/CFT conjecture, where the NHEK  BH can be described by a dual chiral CFT. According to the correspondence, for   every bulk field $\Phi$ there is a corresponding boundary local operator ${\cal O}$ coupled to the boundary field $\Phi_b$ such that 

\begin{equation}
    Z_{\rm AdS,\, eff}[\Phi] = e^{iS_{\rm eff}[\Phi]} = \left \langle {\rm T} \, e^{\int_{\partial \rm AdS} \Phi_b \mathcal{O} } \right \rangle_{\rm CFT},
\end{equation}
where T is the time-ordering operator.
In particular, for the helicity-2 gravitational strain the associated boundary local operator is the stress energy-momentum tensor. The correlators of the QNMs can be therefore inferred from the correlators of the boundary energy-momentum tensor, as we will show in the following.

\vskip 0.5cm
\noindent{{\bf{\it nonlinearities of the QNMs from the Kerr/CFT.}}}
The energy-momentum tensor exists in any local CFT. In two-dimensions, its components are denoted as $T$ and $\overline T$, and they have weights $(h,\bar{h})=(2,0)$  and $(0,2)$, respectively. The central charge $c=12J$ completely fixes their correlators \cite{Guica:2008mu}. 
By defining the energy-momentum tensor as $T_{\mu\nu}$ as the response of the Hamiltonian $\mathcal{H}$ to the transformation $x^\mu \to x^\mu+\xi^\mu$, as 
\begin{equation}
\delta \mathcal{H}=-\int \d^2 x \,\partial_\mu \xi_\nu\, T^{\mu\nu},
\end{equation}
the  two- and three-point correlators  on the complex plane turn out to be \cite{Belavin:1984vu}
\begin{eqnarray}
 &&\langle T(z_1) T(z_2)\rangle=\frac{1}{(2\pi)^2}\frac{c/2}{z_{21}^4}, 
 \label{2pt}\\
 &&\langle T(z_1) T(z_2)T(z_3)\rangle=\frac{1}{(2\pi)^3}\frac{c}{z_{21}^2 \,z_{32}^2\, z_{13}^2}, \label{3pt}
 \end{eqnarray} 
 where $z_{ij}=(z_i-z_j)$ and $T=T_{zz}$.  To take into account that the CFT is at finite temperature $T_L$ for the left-movers, we  map the complex plane to the cylinder with complex coordinate $w=x^1+i \tau$ 
by \cite{DiFrancesco:1997nk}
\begin{eqnarray}
 z=e^{2\pi T_L w}, \qquad \bar z=e^{2\pi T_L \bar{w}}
 \end{eqnarray} 
and identify $\tau$ with $\tau+1/T_L$. Then,  using the transformation property of the energy-momentum tensor under conformal transformations $z\to w(z)$ 
\begin{eqnarray}
T(z)\to T(w)=\left(\partial_z w\right)^2\left(T(z)-\frac{c}{12}
\{w;z\}\right),
\end{eqnarray}
where $\{w;z\}$ is the Schwarzian derivative, we find that the two- and three-point functions at finite temperature $T_L$, after Wick rotating $\tau\to ix^0$, are
\begin{widetext}
\begin{eqnarray}
 &&\langle T(x_1^-) T(x_2^-)\rangle=\frac{c/2}{ (2\pi)^2}\left(\frac{\pi T_L}{\sinh(\pi T_L x_{21}^-)}\right)^4, 
 \label{22pt}\\
 &&\langle T(x_1^-) T(x_2^-)T(x_3^-)\rangle=\frac{c}{(2\pi)^3}\left(\frac{\pi T_L}{\sinh(\pi T_L x_{12}^-)}\right)^2\left(\frac{\pi T_L}{\sinh(\pi T_L x_{23}^-)}\right)^2\left(\frac{\pi T_L}{\sinh(\pi T_L x_{31}^-)}\right)^2, \label{33pt}
 \end{eqnarray} 
\end{widetext}
where $x^-=(x^1-x^0)$. The bulk isometries $\partial_\phi$ and $\partial_t$ are identified, up
to a scale, with the left and right translations in the  CFT, implying $x^-=\phi$ and $x^+=t$.
Going to momentum space and from Eqs. (\ref{ca}) and (\ref{cb}), one identifies therefore the frequency of the left-movers with the azimuthal number $m$ \cite{c1}.

We can write the above connected correlators in momentum space as 
  \begin{align}
 \langle T_{m_1} T_{m_2}\rangle=&\frac{1}{2}
 \int \prod_{i=1}^2 \d x^-_i  
e^{i\o_ix^-_i}\langle T(x^-_1) T(x^-_2)\rangle, \label{2ptk}\nonumber
\end{align}
and
 \begin{align}
\langle T_{\o_1} T_{\o_2}T_{\o_3}\rangle=& 
\int \prod_{i=1}^3 \frac{\d x^-_i}{2^{3/2}}  
e^{i\o_i x^-_i}\langle T(x^-_1) T(x^-_2)T(x^-_3)\rangle, \nonumber
\end{align}
where the $T_m$'s are the boundary duals to the modes of the gravitational strain with angular momentum $m$. 
We obtain 
 \cite{Gubser:1997cm,c1,Becker:2014jla}

\begin{eqnarray}
&&\langle T_{\o_1} T_{\o_2}\rangle'=\frac{c}{24}\frac{(2\pi T_L)^3}{(2\pi)^2}\,
 e^{m_1/(2T_L)} \left|\Gamma\left(2+i\frac{m_1}{2\pi T_L}\right)\right|^2\nonumber
\end{eqnarray}
and 
\begin{widetext}
\begin{eqnarray}
&&\langle T_{\o_1} T_{\o_2}T_{\o_3}\rangle'=- \frac{c}{2\sqrt{2}}\,\frac{(2\pi T_L)^4}{(2\pi)^3}
e^{-(\o_1+\o_2)/(2T_L)}G^{3,3}_{3,3}
\left(\begin{array}{ccc}
-i\frac{\o_1}{2\pi T_L},&0,&i\frac{\o_2}{2\pi T_L}\\
1-i\frac{\o_1}{2\pi T_L},&1,&1+i\frac{\o_2}{2\pi T_L}
\end{array}\Bigg|\,\,  e^{i\pi}\right),
\end{eqnarray}
where the primes indicate we have removed the Dirac delta function $(2\pi)\delta(\sum_i m_i$) and $G^{3,3}_{3,3}$ is a Meijer-$G$ function. 
The  two- and three-point correlators of the graviton are then inferred from  the expressions
\cite{Maldacena:2002vr}

\begin{equation}
\langle \g_{m} h_{-m}\rangle'=-\frac{1}{2\,{\rm Re}\langle T_\o 
T_{-\o}\rangle'}\,,
\quad\quad
\langle \g_{\o_1}\g_{\o_2}\g_{\o_3}\rangle'=\dfrac{2\,{\rm Re}\langle T_{\o_1}T_{\o_2}T_{\o_3}\rangle'}{\prod_{i=1}^3 (-2\, {\rm Re} \langle T_{\o_i} T_{-\o_i}\rangle')  }, \label{3ptg}
\end{equation}
from where it follows that (setting finally  $T_L=1/2\pi$)
\begin{align}
\label{fund}
\frac{\langle \g_{\o_1}\g_{\o_2}
\g_{\o_3}\rangle'}{\langle \g_{\o_1}\g_{-\o_1}\rangle'
\langle \g_{\o_2}\g_{-\o_2}\rangle'}&=-\frac{{\rm Re}\langle T_{\o_1}T_{\o_2}T_{\o_3}\rangle'}{{\rm Re}\langle T_{\o_3} 
T_{-\o_3}\rangle'}=\frac{6\sqrt{2}}{2\pi} \frac{G^{3,3}_{3,3}
\left(\begin{array}{ccc}
-i\o_1,&0,&i\o_2\\
1-i\o_1,&1,&1+i\o_2
\end{array}\Bigg|\,\,  e^{i\pi}\right)}{\left|\Gamma\left(2+im_3\right)\right|^2},
\end{align}
which  critically does not depend upon the central charge~$c$. The last passage is to  integrate over the remaining part of the spin-weighted spherical harmonics in the polar angle~$\theta$. We obtain the general expression

\begin{eqnarray}
\label{resultfund}
\frac{\langle
h_{(\ell_1, m_1)}h_{(\ell_2, m_2)}h_{(\ell_1+\ell_2, m_1+m_2)}
\rangle}{\langle h^2_{(\ell_1, m_1)}\rangle \langle h^2_{(\ell_2, m_2)}\rangle }
&=&\frac{6\sqrt{2}}{2\pi}\, 
{}_{-2} C_{\ell_1, \ell_2,\ell_1+\ell_2}
^{m_1, m_2, m_1+m_2} \,\frac{G^{3,3}_{3,3}
\left(\begin{array}{ccc}
-i\o_1,&0,&i\o_2\\
1-i\o_1,&1,&1+i\o_2
\end{array}\Bigg|\,\,  e^{i\pi}\right)}{\left|\Gamma\left(2-i(m_1+m_2)\right)\right|^2}, 
\end{eqnarray}
where 
\begin{eqnarray}
{}_{-2}C_{\ell_1, \ell_2,\ell_3}
^{m_1, m_2, m_3}&=&2\pi
\int_0^\pi
{\rm d}\theta\,\sin\theta\,
{}_{-2}Y_{(\ell_1, m_1)}{}_{-2}Y_{(\ell_2, m_2)}
{}_{2}\overline{Y}_{(\ell_3, m_3)}\nonumber \\
&=&\frac{\Gamma\bigg(\!\!-2+\sum\limits_{i=1}^3 \displaystyle\frac{|m_i|}{2}\bigg)\Gamma\bigg(4+\sum\limits_{i=1}^3 \displaystyle\frac{|m_i|}{2}\bigg)}{2\sqrt{\pi}\, \Gamma\bigg(2+\sum\limits_{i=1}^3 |m_i|\bigg)}\left(\prod\limits_{i=1}^3\frac{(2|m_i|+1)!}{(|m_i|+2)!\, \,(|m_i|-2)!}\right)^{1/2},
\end{eqnarray}
\end{widetext}
with ${}_{s}\overline{Y}_{(\ell, m)}=(-1)^{s-m}{}_{-s}{Y}_{(\ell, -m)}$,   valid for $\ell_1=m_1$, $\ell_2=m_2$ and $\ell_3=-m_3=(m_1+m_2)$.
Taking $m_1=m_2= 2$, we get\footnote{The fits in Ref. \cite{nl3} are obtained for positive azimuthal numbers. We thank E. Berti and M. Cheung for exchanges about this technical point. However, notice that our results are symmetric under the change of sign of the azimuthal numbers.}



\begin{equation}
\label{result}
\frac{\langle
h_{(2, 2)}h_{(2, 2)}h_{(4, 4)}
\rangle}{\langle h^2_{(2, 2)}\rangle^2 }
\simeq 0.62\cdot\frac{5}{24}\sqrt{\frac{7}{\pi}}\simeq 0.19, 
\end{equation}
which is quite in  good agreement with the numerical results of Refs. \cite{nl2,nl3}. 
To get closer to the  spin value of  $0.7$ considered in those simulations, we can partially approximate departure from extremality using  the corresponding temperature $T_L$ from Eq. (\ref{TLTR}) in our expressions. We obtain in Eq. (\ref{result}) the  numerical value of 0.17,  which is even astonishingly closer to the numerical result in Ref. \cite{nl3}. 
For  the modes  $m_1=2$ and  $m_2=3$, we find (summing up the two terms from the permutation of the modes $m=2$ and 3)
\vspace{-0.2cm}
\begin{equation}
\label{result1}
\frac{\langle
h_{(2, 2)}h_{(3, 3)}h_{(5, 5)}
\rangle}{\langle h^2_{(2, 2)}\rangle
\langle h^2_{(3, 3)}\rangle}
\simeq 1.57\cdot\frac{2}{3}\sqrt{\frac{7}{11\pi}}\simeq 0.47, 
\end{equation}
matching again  the value found in Ref. \cite{nl3} and giving confidence on the validity of 
 Eq. (\ref{resultfund}) (we obtain 0.45 taking into account the  corresponding temperature $T_L$ from Eq. (\ref{TLTR}) for spins equal to 0.7). 
 It would be nice to check the expression (\ref{resultfund})    against numerical quasicircular simulations giving rise to  fast spinning remnants from which  the various multipoles correlators may be extracted.


\vskip 0.5cm
\noindent
\noindent{{\bf{\it Conclusions.}}}
In this paper we have offered an argument based on the Kerr/CFT correspondence to evaluate the nonlinearities of the Kerr BH ringdown. We have shown that the Kerr/CFT correspondence provides a simple way   to both explain, at the qualitative level, the
quadratic scaling of the second-order  mode amplitude with the product of the amplitudes of the fundamental modes, and also to quantitatively predict   the size of this  effect in striking  agreement with  numerical results found in  recent literature. 

Our findings hold only in the extremal case; our next step will be to study departures from this condition. Numerical results are indeed found close to, but not exactly at,  extremality, i.e.  for remnant dimensionless spins $\simeq 0.7$ \cite{nl2,nl3}. Some corrections might therefore intervene.  For  small deviations from extremality, a different  set of boundary conditions  lead to a second copy of the Virasoro algebra \cite{b1} with indications of a hidden dual CFT even far from extremality~\cite{Castro:2010fd,Hui,K}, where the new excitations correspond to right-movers.  However, the fact that our findings are so close to the numerical fits of Refs. \cite{nl2,nl3}  might indicate that the value of the remnant spin is not so relevant (as it also appears from  numerical results) and that the right-mover sector is  decoupled from the dynamics. We have indeed reasons to believe that the right sector
is associated to the  $s=+2$ degree of freedom whose contribution to the GW strain is subleading at infinity with respect to the $s=-2$ mode \cite{inprepP}. Large nonlinearities have been observed also in the case of head-on collisions, giving rise to non-spinning BHs \cite{nl3}. This fact  as well might be explained  using similar symmetry arguments, but in the context of Schwarzschild BHs~\cite{Hui}.
The CFT approach may also provide new consistency relations among the QNM amplitudes and useful predictions for the nonlinearities involved in the QNM dynamics.
Finally, it would  also be interesting to go beyond the three-point correlator, to see if sizeable effects persist at higher-orders. 

We will investigate these issues in  order to provide generic predictions in terms of the multipole numbers and spins of the remnants in  future work~\cite{inprepP}.

\vskip 0.5cm
\noindent
\noindent{{\bf{\it Acknowledgments.}}}
We thank E. Berti for many useful discussions about the results of Ref. \cite{nl3}. We thank  E. Berti, M. Biagetti, and V. De Luca  
for useful feedback on the draft. The work of D. P. and F. R.  is supported by the Swiss National Science Foundation under grants no. 200021-205016 and PP00P2-206149. A.K. is supported by the PEVE-2020 NTUA programme for basic research with project number 65228100. A. K. thanks the University of Geneva where part of this work was performed. A.R. is funded by the Boninchi Foundation.


\bibliographystyle{apsrev4-1}
\bibliography{draft.bib}

\begin{thebibliography}{29}%
\makeatletter
\providecommand \@ifxundefined [1]{%
 \@ifx{#1\undefined}
}%
\providecommand \@ifnum [1]{%
 \ifnum #1\expandafter \@firstoftwo
 \else \expandafter \@secondoftwo
 \fi
}%
\providecommand \@ifx [1]{%
 \ifx #1\expandafter \@firstoftwo
 \else \expandafter \@secondoftwo
 \fi
}%
\providecommand \natexlab [1]{#1}%
\providecommand \enquote  [1]{``#1''}%
\providecommand \bibnamefont  [1]{#1}%
\providecommand \bibfnamefont [1]{#1}%
\providecommand \citenamefont [1]{#1}%
\providecommand \href@noop [0]{\@secondoftwo}%
\providecommand \href [0]{\begingroup \@sanitize@url \@href}%
\providecommand \@href[1]{\@@startlink{#1}\@@href}%
\providecommand \@@href[1]{\endgroup#1\@@endlink}%
\providecommand \@sanitize@url [0]{\catcode `\\12\catcode `\$12\catcode
  `\&12\catcode `\#12\catcode `\^12\catcode `\_12\catcode `\%12\relax}%
\providecommand \@@startlink[1]{}%
\providecommand \@@endlink[0]{}%
\providecommand \url  [0]{\begingroup\@sanitize@url \@url }%
\providecommand \@url [1]{\endgroup\@href {#1}{\urlprefix }}%
\providecommand \urlprefix  [0]{URL }%
\providecommand \Eprint [0]{\href }%
\providecommand \doibase [0]{http://dx.doi.org/}%
\providecommand \selectlanguage [0]{\@gobble}%
\providecommand \bibinfo  [0]{\@secondoftwo}%
\providecommand \bibfield  [0]{\@secondoftwo}%
\providecommand \translation [1]{[#1]}%
\providecommand \BibitemOpen [0]{}%
\providecommand \bibitemStop [0]{}%
\providecommand \bibitemNoStop [0]{.\EOS\space}%
\providecommand \EOS [0]{\spacefactor3000\relax}%
\providecommand \BibitemShut  [1]{\csname bibitem#1\endcsname}%
\let\auto@bib@innerbib\@empty
\bibitem [{\citenamefont {Berti}\ \emph {et~al.}(2009)\citenamefont {Berti},
  \citenamefont {Cardoso},\ and\ \citenamefont {Starinets}}]{Berti:2009kk}%
  \BibitemOpen
  \bibfield  {author} {\bibinfo {author} {\bibfnamefont {E.}~\bibnamefont
  {Berti}}, \bibinfo {author} {\bibfnamefont {V.}~\bibnamefont {Cardoso}}, \
  and\ \bibinfo {author} {\bibfnamefont {A.~O.}\ \bibnamefont {Starinets}},\
  }\href {\doibase 10.1088/0264-9381/26/16/163001} {\bibfield  {journal}
  {\bibinfo  {journal} {Class. Quant. Grav.}\ }\textbf {\bibinfo {volume}
  {26}},\ \bibinfo {pages} {163001} (\bibinfo {year} {2009})},\ \Eprint
  {http://arxiv.org/abs/0905.2975} {arXiv:0905.2975 [gr-qc]} \BibitemShut
  {NoStop}%
\bibitem [{\citenamefont {Teukolsky}(1973)}]{Teukolsky:1973ha}%
  \BibitemOpen
  \bibfield  {author} {\bibinfo {author} {\bibfnamefont {S.~A.}\ \bibnamefont
  {Teukolsky}},\ }\href {\doibase 10.1086/152444} {\bibfield  {journal}
  {\bibinfo  {journal} {Astrophys. J.}\ }\textbf {\bibinfo {volume} {185}},\
  \bibinfo {pages} {635} (\bibinfo {year} {1973})}\BibitemShut {NoStop}%
\bibitem [{\citenamefont {Cheung}\ \emph {et~al.}(2022)\citenamefont {Cheung}
  \emph {et~al.}}]{nl3}%
  \BibitemOpen
  \bibfield  {author} {\bibinfo {author} {\bibfnamefont {M.~H.-Y.}\
  \bibnamefont {Cheung}} \emph {et~al.},\ }\href@noop {} {\  (\bibinfo {year}
  {2022})},\ \Eprint {http://arxiv.org/abs/2208.07374} {arXiv:2208.07374
  [gr-qc]} \BibitemShut {NoStop}%
\bibitem [{\citenamefont {Mitman}\ \emph {et~al.}(2022)\citenamefont {Mitman}
  \emph {et~al.}}]{nl2}%
  \BibitemOpen
  \bibfield  {author} {\bibinfo {author} {\bibfnamefont {K.}~\bibnamefont
  {Mitman}} \emph {et~al.},\ }\href@noop {} {\  (\bibinfo {year} {2022})},\
  \Eprint {http://arxiv.org/abs/2208.07380} {arXiv:2208.07380 [gr-qc]}
  \BibitemShut {NoStop}%
\bibitem [{\citenamefont {London}\ \emph {et~al.}(2014)\citenamefont {London},
  \citenamefont {Shoemaker},\ and\ \citenamefont {Healy}}]{nlold1}%
  \BibitemOpen
  \bibfield  {author} {\bibinfo {author} {\bibfnamefont {L.}~\bibnamefont
  {London}}, \bibinfo {author} {\bibfnamefont {D.}~\bibnamefont {Shoemaker}}, \
  and\ \bibinfo {author} {\bibfnamefont {J.}~\bibnamefont {Healy}},\ }\href
  {\doibase 10.1103/PhysRevD.90.124032} {\bibfield  {journal} {\bibinfo
  {journal} {Phys. Rev. D}\ }\textbf {\bibinfo {volume} {90}},\ \bibinfo
  {pages} {124032} (\bibinfo {year} {2014})},\ \bibinfo {note} {[Erratum:
  Phys.Rev.D 94, 069902 (2016)]},\ \Eprint {http://arxiv.org/abs/1404.3197}
  {arXiv:1404.3197 [gr-qc]} \BibitemShut {NoStop}%
\bibitem [{\citenamefont {Ma}\ \emph {et~al.}(2022)\citenamefont {Ma},
  \citenamefont {Mitman}, \citenamefont {Sun}, \citenamefont {Deppe},
  \citenamefont {H\'ebert}, \citenamefont {Kidder}, \citenamefont {Moxon},
  \citenamefont {Throwe}, \citenamefont {Vu},\ and\ \citenamefont
  {Chen}}]{nlold2}%
  \BibitemOpen
  \bibfield  {author} {\bibinfo {author} {\bibfnamefont {S.}~\bibnamefont
  {Ma}}, \bibinfo {author} {\bibfnamefont {K.}~\bibnamefont {Mitman}}, \bibinfo
  {author} {\bibfnamefont {L.}~\bibnamefont {Sun}}, \bibinfo {author}
  {\bibfnamefont {N.}~\bibnamefont {Deppe}}, \bibinfo {author} {\bibfnamefont
  {F.}~\bibnamefont {H\'ebert}}, \bibinfo {author} {\bibfnamefont {L.~E.}\
  \bibnamefont {Kidder}}, \bibinfo {author} {\bibfnamefont {J.}~\bibnamefont
  {Moxon}}, \bibinfo {author} {\bibfnamefont {W.}~\bibnamefont {Throwe}},
  \bibinfo {author} {\bibfnamefont {N.~L.}\ \bibnamefont {Vu}}, \ and\ \bibinfo
  {author} {\bibfnamefont {Y.}~\bibnamefont {Chen}},\ }\href {\doibase
  10.1103/PhysRevD.106.084036} {\bibfield  {journal} {\bibinfo  {journal}
  {Phys. Rev. D}\ }\textbf {\bibinfo {volume} {106}},\ \bibinfo {pages}
  {084036} (\bibinfo {year} {2022})},\ \Eprint
  {http://arxiv.org/abs/2207.10870} {arXiv:2207.10870 [gr-qc]} \BibitemShut
  {NoStop}%
\bibitem [{\citenamefont {Lagos}\ and\ \citenamefont
  {Hui}(2022)}]{Lagos:2022otp}%
  \BibitemOpen
  \bibfield  {author} {\bibinfo {author} {\bibfnamefont {M.}~\bibnamefont
  {Lagos}}\ and\ \bibinfo {author} {\bibfnamefont {L.}~\bibnamefont {Hui}},\
  }\href@noop {} {\  (\bibinfo {year} {2022})},\ \Eprint
  {http://arxiv.org/abs/2208.07379} {arXiv:2208.07379 [gr-qc]} \BibitemShut
  {NoStop}%
\bibitem [{\citenamefont {Maggiore}(2018)}]{Maggiore:2018sht}%
  \BibitemOpen
  \bibfield  {author} {\bibinfo {author} {\bibfnamefont {M.}~\bibnamefont
  {Maggiore}},\ }\href@noop {} {\emph {\bibinfo {title} {{Gravitational Waves.
  Vol. 2: Astrophysics and Cosmology}}}}\ (\bibinfo  {publisher} {Oxford
  University Press},\ \bibinfo {year} {2018})\BibitemShut {NoStop}%
\bibitem [{\citenamefont {Guica}\ \emph {et~al.}(2009)\citenamefont {Guica},
  \citenamefont {Hartman}, \citenamefont {Song},\ and\ \citenamefont
  {Strominger}}]{Guica:2008mu}%
  \BibitemOpen
  \bibfield  {author} {\bibinfo {author} {\bibfnamefont {M.}~\bibnamefont
  {Guica}}, \bibinfo {author} {\bibfnamefont {T.}~\bibnamefont {Hartman}},
  \bibinfo {author} {\bibfnamefont {W.}~\bibnamefont {Song}}, \ and\ \bibinfo
  {author} {\bibfnamefont {A.}~\bibnamefont {Strominger}},\ }\href {\doibase
  10.1103/PhysRevD.80.124008} {\bibfield  {journal} {\bibinfo  {journal} {Phys.
  Rev. D}\ }\textbf {\bibinfo {volume} {80}},\ \bibinfo {pages} {124008}
  (\bibinfo {year} {2009})},\ \Eprint {http://arxiv.org/abs/0809.4266}
  {arXiv:0809.4266 [hep-th]} \BibitemShut {NoStop}%
\bibitem [{\citenamefont {Bredberg}\ \emph {et~al.}(2010)\citenamefont
  {Bredberg}, \citenamefont {Hartman}, \citenamefont {Song},\ and\
  \citenamefont {Strominger}}]{c1}%
  \BibitemOpen
  \bibfield  {author} {\bibinfo {author} {\bibfnamefont {I.}~\bibnamefont
  {Bredberg}}, \bibinfo {author} {\bibfnamefont {T.}~\bibnamefont {Hartman}},
  \bibinfo {author} {\bibfnamefont {W.}~\bibnamefont {Song}}, \ and\ \bibinfo
  {author} {\bibfnamefont {A.}~\bibnamefont {Strominger}},\ }\href {\doibase
  10.1007/JHEP04(2010)019} {\bibfield  {journal} {\bibinfo  {journal} {JHEP}\
  }\textbf {\bibinfo {volume} {04}},\ \bibinfo {pages} {019} (\bibinfo {year}
  {2010})},\ \Eprint {http://arxiv.org/abs/0907.3477} {arXiv:0907.3477
  [hep-th]} \BibitemShut {NoStop}%
\bibitem [{\citenamefont {Hartman}\ \emph {et~al.}(2010)\citenamefont
  {Hartman}, \citenamefont {Song},\ and\ \citenamefont {Strominger}}]{c2}%
  \BibitemOpen
  \bibfield  {author} {\bibinfo {author} {\bibfnamefont {T.}~\bibnamefont
  {Hartman}}, \bibinfo {author} {\bibfnamefont {W.}~\bibnamefont {Song}}, \
  and\ \bibinfo {author} {\bibfnamefont {A.}~\bibnamefont {Strominger}},\
  }\href {\doibase 10.1007/JHEP03(2010)118} {\bibfield  {journal} {\bibinfo
  {journal} {JHEP}\ }\textbf {\bibinfo {volume} {03}},\ \bibinfo {pages} {118}
  (\bibinfo {year} {2010})},\ \Eprint {http://arxiv.org/abs/0908.3909}
  {arXiv:0908.3909 [hep-th]} \BibitemShut {NoStop}%
\bibitem [{\citenamefont {Porfyriadis}\ and\ \citenamefont
  {Strominger}(2014)}]{c3}%
  \BibitemOpen
  \bibfield  {author} {\bibinfo {author} {\bibfnamefont {A.~P.}\ \bibnamefont
  {Porfyriadis}}\ and\ \bibinfo {author} {\bibfnamefont {A.}~\bibnamefont
  {Strominger}},\ }\href {\doibase 10.1103/PhysRevD.90.044038} {\bibfield
  {journal} {\bibinfo  {journal} {Phys. Rev. D}\ }\textbf {\bibinfo {volume}
  {90}},\ \bibinfo {pages} {044038} (\bibinfo {year} {2014})},\ \Eprint
  {http://arxiv.org/abs/1401.3746} {arXiv:1401.3746 [hep-th]} \BibitemShut
  {NoStop}%
\bibitem [{\citenamefont {Castro}\ and\ \citenamefont {Larsen}(2009)}]{b1}%
  \BibitemOpen
  \bibfield  {author} {\bibinfo {author} {\bibfnamefont {A.}~\bibnamefont
  {Castro}}\ and\ \bibinfo {author} {\bibfnamefont {F.}~\bibnamefont
  {Larsen}},\ }\href {\doibase 10.1088/1126-6708/2009/12/037} {\bibfield
  {journal} {\bibinfo  {journal} {JHEP}\ }\textbf {\bibinfo {volume} {12}},\
  \bibinfo {pages} {037} (\bibinfo {year} {2009})},\ \Eprint
  {http://arxiv.org/abs/0908.1121} {arXiv:0908.1121 [hep-th]} \BibitemShut
  {NoStop}%
\bibitem [{\citenamefont {Matsuo}\ \emph {et~al.}(2010)\citenamefont {Matsuo},
  \citenamefont {Tsukioka},\ and\ \citenamefont {Yoo}}]{b2}%
  \BibitemOpen
  \bibfield  {author} {\bibinfo {author} {\bibfnamefont {Y.}~\bibnamefont
  {Matsuo}}, \bibinfo {author} {\bibfnamefont {T.}~\bibnamefont {Tsukioka}}, \
  and\ \bibinfo {author} {\bibfnamefont {C.-M.}\ \bibnamefont {Yoo}},\ }\href
  {\doibase 10.1016/j.nuclphysb.2009.09.025} {\bibfield  {journal} {\bibinfo
  {journal} {Nucl. Phys. B}\ }\textbf {\bibinfo {volume} {825}},\ \bibinfo
  {pages} {231} (\bibinfo {year} {2010})},\ \Eprint
  {http://arxiv.org/abs/0907.0303} {arXiv:0907.0303 [hep-th]} \BibitemShut
  {NoStop}%
\bibitem [{\citenamefont {Rasmussen}(2010)}]{b3}%
  \BibitemOpen
  \bibfield  {author} {\bibinfo {author} {\bibfnamefont {J.}~\bibnamefont
  {Rasmussen}},\ }\href {\doibase 10.1142/S0217751X10048986} {\bibfield
  {journal} {\bibinfo  {journal} {Int. J. Mod. Phys. A}\ }\textbf {\bibinfo
  {volume} {25}},\ \bibinfo {pages} {1597} (\bibinfo {year} {2010})},\ \Eprint
  {http://arxiv.org/abs/0908.0184} {arXiv:0908.0184 [hep-th]} \BibitemShut
  {NoStop}%
\bibitem [{\citenamefont {Becker}\ \emph {et~al.}(2011)\citenamefont {Becker},
  \citenamefont {Cremonini},\ and\ \citenamefont {Schulgin}}]{Becker:2010jj}%
  \BibitemOpen
  \bibfield  {author} {\bibinfo {author} {\bibfnamefont {M.}~\bibnamefont
  {Becker}}, \bibinfo {author} {\bibfnamefont {S.}~\bibnamefont {Cremonini}}, \
  and\ \bibinfo {author} {\bibfnamefont {W.}~\bibnamefont {Schulgin}},\ }\href
  {\doibase 10.1007/JHEP02(2011)007} {\bibfield  {journal} {\bibinfo  {journal}
  {JHEP}\ }\textbf {\bibinfo {volume} {02}},\ \bibinfo {pages} {007} (\bibinfo
  {year} {2011})},\ \Eprint {http://arxiv.org/abs/1004.1174} {arXiv:1004.1174
  [hep-th]} \BibitemShut {NoStop}%
\bibitem [{\citenamefont {Bardeen}\ and\ \citenamefont
  {Horowitz}(1999)}]{Bardeen:1999px}%
  \BibitemOpen
  \bibfield  {author} {\bibinfo {author} {\bibfnamefont {J.~M.}\ \bibnamefont
  {Bardeen}}\ and\ \bibinfo {author} {\bibfnamefont {G.~T.}\ \bibnamefont
  {Horowitz}},\ }\href {\doibase 10.1103/PhysRevD.60.104030} {\bibfield
  {journal} {\bibinfo  {journal} {Phys. Rev. D}\ }\textbf {\bibinfo {volume}
  {60}},\ \bibinfo {pages} {104030} (\bibinfo {year} {1999})},\ \Eprint
  {http://arxiv.org/abs/hep-th/9905099} {arXiv:hep-th/9905099} \BibitemShut
  {NoStop}%
\bibitem [{\citenamefont {Anninos}\ \emph {et~al.}(2009)\citenamefont
  {Anninos}, \citenamefont {Li}, \citenamefont {Padi}, \citenamefont {Song},\
  and\ \citenamefont {Strominger}}]{Anninos:2008fx}%
  \BibitemOpen
  \bibfield  {author} {\bibinfo {author} {\bibfnamefont {D.}~\bibnamefont
  {Anninos}}, \bibinfo {author} {\bibfnamefont {W.}~\bibnamefont {Li}},
  \bibinfo {author} {\bibfnamefont {M.}~\bibnamefont {Padi}}, \bibinfo {author}
  {\bibfnamefont {W.}~\bibnamefont {Song}}, \ and\ \bibinfo {author}
  {\bibfnamefont {A.}~\bibnamefont {Strominger}},\ }\href {\doibase
  10.1088/1126-6708/2009/03/130} {\bibfield  {journal} {\bibinfo  {journal}
  {JHEP}\ }\textbf {\bibinfo {volume} {03}},\ \bibinfo {pages} {130} (\bibinfo
  {year} {2009})},\ \Eprint {http://arxiv.org/abs/0807.3040} {arXiv:0807.3040
  [hep-th]} \BibitemShut {NoStop}%
\bibitem [{\citenamefont {Bengtsson}\ and\ \citenamefont
  {Sandin}(2006)}]{Bengtsson:2005zj}%
  \BibitemOpen
  \bibfield  {author} {\bibinfo {author} {\bibfnamefont {I.}~\bibnamefont
  {Bengtsson}}\ and\ \bibinfo {author} {\bibfnamefont {P.}~\bibnamefont
  {Sandin}},\ }\href {\doibase 10.1088/0264-9381/23/3/022} {\bibfield
  {journal} {\bibinfo  {journal} {Class. Quant. Grav.}\ }\textbf {\bibinfo
  {volume} {23}},\ \bibinfo {pages} {971} (\bibinfo {year} {2006})},\ \Eprint
  {http://arxiv.org/abs/gr-qc/0509076} {arXiv:gr-qc/0509076} \BibitemShut
  {NoStop}%
\bibitem [{\citenamefont {Frolov}\ and\ \citenamefont
  {Thorne}(1989)}]{Frolov:1989jh}%
  \BibitemOpen
  \bibfield  {author} {\bibinfo {author} {\bibfnamefont {V.~P.}\ \bibnamefont
  {Frolov}}\ and\ \bibinfo {author} {\bibfnamefont {K.~S.}\ \bibnamefont
  {Thorne}},\ }\href {\doibase 10.1103/PhysRevD.39.2125} {\bibfield  {journal}
  {\bibinfo  {journal} {Phys. Rev. D}\ }\textbf {\bibinfo {volume} {39}},\
  \bibinfo {pages} {2125} (\bibinfo {year} {1989})}\BibitemShut {NoStop}%
\bibitem [{\citenamefont {Belavin}\ \emph {et~al.}(1984)\citenamefont
  {Belavin}, \citenamefont {Polyakov},\ and\ \citenamefont
  {Zamolodchikov}}]{Belavin:1984vu}%
  \BibitemOpen
  \bibfield  {author} {\bibinfo {author} {\bibfnamefont {A.~A.}\ \bibnamefont
  {Belavin}}, \bibinfo {author} {\bibfnamefont {A.~M.}\ \bibnamefont
  {Polyakov}}, \ and\ \bibinfo {author} {\bibfnamefont {A.~B.}\ \bibnamefont
  {Zamolodchikov}},\ }\href {\doibase 10.1016/0550-3213(84)90052-X} {\bibfield
  {journal} {\bibinfo  {journal} {Nucl. Phys. B}\ }\textbf {\bibinfo {volume}
  {241}},\ \bibinfo {pages} {333} (\bibinfo {year} {1984})}\BibitemShut
  {NoStop}%
\bibitem [{\citenamefont {Di~Francesco}\ \emph {et~al.}(1997)\citenamefont
  {Di~Francesco}, \citenamefont {Mathieu},\ and\ \citenamefont
  {Senechal}}]{DiFrancesco:1997nk}%
  \BibitemOpen
  \bibfield  {author} {\bibinfo {author} {\bibfnamefont {P.}~\bibnamefont
  {Di~Francesco}}, \bibinfo {author} {\bibfnamefont {P.}~\bibnamefont
  {Mathieu}}, \ and\ \bibinfo {author} {\bibfnamefont {D.}~\bibnamefont
  {Senechal}},\ }\href {\doibase 10.1007/978-1-4612-2256-9} {\emph {\bibinfo
  {title} {{Conformal Field Theory}}}},\ Graduate Texts in Contemporary
  Physics\ (\bibinfo  {publisher} {Springer-Verlag},\ \bibinfo {address} {New
  York},\ \bibinfo {year} {1997})\BibitemShut {NoStop}%
\bibitem [{\citenamefont {Gubser}(1997)}]{Gubser:1997cm}%
  \BibitemOpen
  \bibfield  {author} {\bibinfo {author} {\bibfnamefont {S.~S.}\ \bibnamefont
  {Gubser}},\ }\href {\doibase 10.1103/PhysRevD.56.7854} {\bibfield  {journal}
  {\bibinfo  {journal} {Phys. Rev. D}\ }\textbf {\bibinfo {volume} {56}},\
  \bibinfo {pages} {7854} (\bibinfo {year} {1997})},\ \Eprint
  {http://arxiv.org/abs/hep-th/9706100} {arXiv:hep-th/9706100} \BibitemShut
  {NoStop}%
\bibitem [{\citenamefont {Becker}\ \emph {et~al.}(2014)\citenamefont {Becker},
  \citenamefont {Cabrera},\ and\ \citenamefont {Su}}]{Becker:2014jla}%
  \BibitemOpen
  \bibfield  {author} {\bibinfo {author} {\bibfnamefont {M.}~\bibnamefont
  {Becker}}, \bibinfo {author} {\bibfnamefont {Y.}~\bibnamefont {Cabrera}}, \
  and\ \bibinfo {author} {\bibfnamefont {N.}~\bibnamefont {Su}},\ }\href
  {\doibase 10.1007/JHEP09(2014)157} {\bibfield  {journal} {\bibinfo  {journal}
  {JHEP}\ }\textbf {\bibinfo {volume} {09}},\ \bibinfo {pages} {157} (\bibinfo
  {year} {2014})},\ \Eprint {http://arxiv.org/abs/1407.3415} {arXiv:1407.3415
  [hep-th]} \BibitemShut {NoStop}%
\bibitem [{\citenamefont {Maldacena}(2003)}]{Maldacena:2002vr}%
  \BibitemOpen
  \bibfield  {author} {\bibinfo {author} {\bibfnamefont {J.~M.}\ \bibnamefont
  {Maldacena}},\ }\href {\doibase 10.1088/1126-6708/2003/05/013} {\bibfield
  {journal} {\bibinfo  {journal} {JHEP}\ }\textbf {\bibinfo {volume} {05}},\
  \bibinfo {pages} {013} (\bibinfo {year} {2003})},\ \Eprint
  {http://arxiv.org/abs/astro-ph/0210603} {arXiv:astro-ph/0210603} \BibitemShut
  {NoStop}%
\bibitem [{\citenamefont {Castro}\ \emph {et~al.}(2010)\citenamefont {Castro},
  \citenamefont {Maloney},\ and\ \citenamefont {Strominger}}]{Castro:2010fd}%
  \BibitemOpen
  \bibfield  {author} {\bibinfo {author} {\bibfnamefont {A.}~\bibnamefont
  {Castro}}, \bibinfo {author} {\bibfnamefont {A.}~\bibnamefont {Maloney}}, \
  and\ \bibinfo {author} {\bibfnamefont {A.}~\bibnamefont {Strominger}},\
  }\href {\doibase 10.1103/PhysRevD.82.024008} {\bibfield  {journal} {\bibinfo
  {journal} {Phys. Rev. D}\ }\textbf {\bibinfo {volume} {82}},\ \bibinfo
  {pages} {024008} (\bibinfo {year} {2010})},\ \Eprint
  {http://arxiv.org/abs/1004.0996} {arXiv:1004.0996 [hep-th]} \BibitemShut
  {NoStop}%
\bibitem [{\citenamefont {Hui}\ \emph {et~al.}(2022)\citenamefont {Hui},
  \citenamefont {Joyce}, \citenamefont {Penco}, \citenamefont {Santoni},\ and\
  \citenamefont {Solomon}}]{Hui}%
  \BibitemOpen
  \bibfield  {author} {\bibinfo {author} {\bibfnamefont {L.}~\bibnamefont
  {Hui}}, \bibinfo {author} {\bibfnamefont {A.}~\bibnamefont {Joyce}}, \bibinfo
  {author} {\bibfnamefont {R.}~\bibnamefont {Penco}}, \bibinfo {author}
  {\bibfnamefont {L.}~\bibnamefont {Santoni}}, \ and\ \bibinfo {author}
  {\bibfnamefont {A.~R.}\ \bibnamefont {Solomon}},\ }\href {\doibase
  10.1007/JHEP09(2022)049} {\bibfield  {journal} {\bibinfo  {journal} {JHEP}\
  }\textbf {\bibinfo {volume} {09}},\ \bibinfo {pages} {049} (\bibinfo {year}
  {2022})},\ \Eprint {http://arxiv.org/abs/2203.08832} {arXiv:2203.08832
  [hep-th]} \BibitemShut {NoStop}%
\bibitem [{\citenamefont {Kehagias}\ \emph {et~al.}(2022)\citenamefont
  {Kehagias}, \citenamefont {Perrone},\ and\ \citenamefont {Riotto}}]{K}%
  \BibitemOpen
  \bibfield  {author} {\bibinfo {author} {\bibfnamefont {A.}~\bibnamefont
  {Kehagias}}, \bibinfo {author} {\bibfnamefont {D.}~\bibnamefont {Perrone}}, \
  and\ \bibinfo {author} {\bibfnamefont {A.}~\bibnamefont {Riotto}},\
  }\href@noop {} {\  (\bibinfo {year} {2022})},\ \Eprint
  {http://arxiv.org/abs/2211.02384} {arXiv:2211.02384 [hep-th]} \BibitemShut
  {NoStop}%
\bibitem [{\citenamefont {Kehagias}\ \emph {et~al.}()\citenamefont {Kehagias},
  \citenamefont {Perrone}, \citenamefont {Riotto},\ and\ \citenamefont
  {Riva}}]{inprepP}%
  \BibitemOpen
  \bibfield  {author} {\bibinfo {author} {\bibfnamefont {A.}~\bibnamefont
  {Kehagias}}, \bibinfo {author} {\bibfnamefont {D.}~\bibnamefont {Perrone}},
  \bibinfo {author} {\bibfnamefont {A.}~\bibnamefont {Riotto}}, \ and\ \bibinfo
  {author} {\bibfnamefont {F.}~\bibnamefont {Riva}},\ }\href@noop {} {\
  }\bibinfo {note} {{in preparation}}\BibitemShut {NoStop}%
\end{thebibliography}%

\end{document}